\documentclass[11pt,aps,pra]{revtex4-1}
\usepackage{amssymb,amsmath}
\usepackage{graphicx}
\usepackage[matrix,frame,arrow]{xy}%    Q-circuit version 1.2
\usepackage[matrix,frame,arrow]{xy}
\usepackage{amsmath}

\newcommand{\qw}[1][-1]{\ar @{-} [0,#1]}
\newcommand{\gate}[1]{*{\xy *+<.6em>{#1};p\save+LU;+RU **\dir{-}\restore\save+RU;+RD **\dir{-}\restore\save+RD;+LD **\dir{-}\restore\POS+LD;+LU **\dir{-}\endxy} \qw}
\newcommand{\measureD}[1]{*{\xy*+=+<.5em>{\vphantom{\rule{0em}{.1em}#1}}*\cir{r_l};p\save*!R{#1} \restore\save+UC;+UC-<.5em,0em>*!R{\hphantom{#1}}+L **\dir{-} \restore\save+DC;+DC-<.5em,0em>*!R{\hphantom{#1}}+L **\dir{-} \restore\POS+UC-<.5em,0em>*!R{\hphantom{#1}}+L;+DC-<.5em,0em>*!R{\hphantom{#1}}+L **\dir{-} \endxy} \qw}
\newcommand{\multigate}[2]{*+<1em,.9em>{\hphantom{#2}} \qw \POS[0,0].[#1,0];p !C *{#2},p \save+LU;+RU **\dir{-}\restore\save+RU;+RD **\dir{-}\restore\save+RD;+LD **\dir{-}\restore\save+LD;+LU **\dir{-}\restore}
\newcommand{\ghost}[1]{*+<1em,.9em>{\hphantom{#1}} \qw}

\newcommand{\Qcircuit}[1][0em]{\xymatrix @*[o] @*=<#1>}  %tentativo disperato
 \renewcommand{\Qcircuit}[1][0em]{\xymatrix @*=<#1>}

\newcommand{\prepareC}[1]{*{\xy*+=+<.5em>{\vphantom{#1\rule{0em}{.1em}}}*\cir{l^r};p\save*!L{#1} \restore\save+UC;+UC+<.5em,0em>*!L{\hphantom{#1}}+R **\dir{-} \restore\save+DC;+DC+<.5em,0em>*!L{\hphantom{#1}}+R **\dir{-} \restore\POS+UC+<.5em,0em>*!L{\hphantom{#1}}+R;+DC+<.5em,0em>*!L{\hphantom{#1}}+R **\dir{-} \endxy}}
\newcommand{\poloFantasmaCn}[1]{{{}^{#1}_{\phantom{#1}}}}

\newtheorem{prin}{Principle}

\begin{document}

\title{Quantum Theory, Namely the Pure and Reversible Theory of Information}

\author{Giulio Chiribella}\email{gchiribella@mail.tsinghua.edu.cn} 
\affiliation{Center for Quantum Information, Institute for Interdisciplinary Information Sciences, Tsinghua University, Beijing, China}
\homepage{http://iiis.tsinghua.edu.cn/giulio/}
\author{Giacomo Mauro D'Ariano}\email{dariano@unipv.it}
\affiliation{{\em QUIT Group}, Dipartimento di Fisica, via Bassi 6, 27100 Pavia, Italy}
\homepage{http://www.qubit.it}
\author{Paolo Perinotti}\email{paolo.perinotti@unipv.it} 
\affiliation{{\em QUIT Group}, Dipartimento di Fisica, via Bassi 6, 27100 Pavia, Italy}
\homepage{http://www.qubit.it}
\date{\today}

% Abstract

%\abstract{After more than a century since its birth, Quantum Theory
%  still eludes our understanding. If asked to describe it, we have to
 % resort to abstract and \emph{ad hoc} principles about complex
 % Hilbert spaces. How is it possible that a fundamental physical
 % theory cannot be described using the ordinary language of Physics?
 % Here we offer a contribution to the problem from the angle of
 % Quantum Information, providing a short non-technical presentation of
 % a recent derivation of Quantum Theory from information-theoretic
 % principles \cite{qmfrompuri}. The broad picture emerging from the
 % principles is that Quantum Theory is the only standard theory of
  %information compatible with the purity and reversibility of physical
  %processes.}

\begin{abstract}
After more than a century since its birth, Quantum Theory
  still eludes our understanding. If asked to describe it, we have to
  resort to abstract and \emph{ad hoc} principles about complex
  Hilbert spaces. How is it possible that a fundamental physical
  theory cannot be described using the ordinary language of Physics?
  Here we offer a contribution to the problem from the angle of
  Quantum Information, providing a short non-technical presentation of
  a recent derivation of Quantum Theory from information-theoretic
  principles \cite{qmfrompuri}. The broad picture emerging from the
  principles is that Quantum Theory is the only standard theory of
  information compatible with the purity and reversibility of physical
  processes.
  \end{abstract}

\maketitle  

% Keywords: add 3 to 10 keywords

%\keyword{Foundations of Quantum Mechanics; Quantum Information; Purification}

% the fields PACS and MSC may be left empty or commented out if not applicable
%\PACS{}
%\MSC{}

%\begin{document}
%%%%%%%%%%%%%%%%%%%%%%%%%%%%%%%%%%%%%%%%%%%%%%%%%%%%%%%%%%%%

\section{Introduction}

Quantum Theory is booming: It allows us to describe elementary
particles and fundamental forces, to predict the colour of the light
emitted by excited atoms and molecules, to explain the black body
spectrum and the photoelectric effect, to determine the specific heat
and the speed of sound in solids, to understand chemical and
biochemical reactions, to construct lasers, transistors, and
computers. This extraordinary experimental and technological success,
however, is dimmed by huge conceptual difficulties. After more than
hundred years from the birth of Quantum Theory, we still struggle to
understand its puzzles and hotly debate on its interpretations. And
even leaving aside the vexed issue of interpretations, there is a more
basic (and embarrassing) problem: We cannot even tell what Quantum
Theory is without resorting to the abstract language of Hilbert
spaces! Compare quantum mechanics with the classical mechanics of
Newton and Laplace: Intuitive notions, such as position and velocity
of a particle, are now replaced by abstract ones, such as unit vector
in a complex Hilbert space. Physical systems are now represented by
Hilbert spaces, pure states by unit vectors, and physical quantities
by self-adjoint operators. What does this mean? Why should Nature be
described by this very special piece of mathematics?

It is hard not to suspect that, despite all our experimental and
technological advancement, we are completely missing the big picture.
The situation was vividly portrayed by John Wheeler in a popular article on the New York Times, where he tried to attract the attention of the general public to what he was considering  ``the greatest mystery in physics today" \cite{wheelerNYT}:
``Balancing the glory of quantum achievements, we have the shame of
not knowing ``how come." Why does the quantum exist?"

The need for a more fundamental understanding was clear since the
early days of Quantum Theory.  The first to be dissatisfied with the
Hilbert space formulation was its founder himself, John von Neumann
\cite{redei}.  Few years after the completion of his monumental book
\cite{von32}, von Neumann tried to understand Quantum Theory as a new
form of logics.  His seminal work in collaboration with Birkhoff
\cite{BirkVN36} originated the field of quantum logics, which however
did not succeed in producing a clear-cut picture capable to cross the
borders of a small community of specialists.  More recently, a fresh
perspective on the origin of the quantum came from Wheeler.  In his
programme \emph{It from Bit}, Wheeler argued that information should
be the fundamental notion in our understanding of the whole of
physics, based on the premise that ``all things physical are information-theoretic in origin''
\cite{wheelerIt}.  If we accept this premise, then nothing is more
natural then looking for an information-theoretic understanding of
\emph{quantum} physics.  Indeed, one of the most noteworthy features
of quantum theory is the peculiar way in which it describes the
extraction of information through measurements.  This remarkable feature and its foundational import were discussed
in depth by Wootters in his PhD thesis \cite{woot}.   In different guises, the idea of
information being the core of Quantum Theory has been explored by
several authors, notably by Weizsacker \cite{weiz}, Zeilinger
\cite{zei}, and Brukner \cite{bruzei}). 

% {\bf  the promise of quantum information, the long quest} 
The idea that Quantum Theory is, in its backbone, a new theory of
information became very concrete with the raise of Quantum
Information.  This revolutionary discipline revealed that Quantum
Theory is not just a theory of unavoidable indeterminacy, as
emphasized by its founders, but also a theory of new exciting ways to
process information, ways that were unimaginable in the old classical
world of Newton and Laplace.  Quantum Information unearthed a huge
number of operational consequences of Quantum Theory: quantum states
cannot be copied \cite{wootterszurek,dieks} but they can be teleported
\cite{tele}, the quantum laws allow for secure key distribution
\cite{bb84,e91}, for fast database search \cite{grover}, and for the
factorization of large numbers in polynomial time \cite{shor}. These
facts are so impressive that one may be tempted to promote some of
them to the role of fundamental principles, trying to derive the
obscure mathematics of Quantum Theory from them. The idea that the
new discoveries of Quantum Information could offer the key to the mystery of
the quantum was enthusiastically championed by Fuchs \cite{fuchs} and
Brassard \cite{brassard} and rapidly led to a feverish quest for new
information-theoretic principles, like \emph{information causality}
\cite{infocau}, and to reconstructions of quantum theory from various
informational ideas, like those of Refs.
\cite{Har01,maurolast,philip,DakBru09,Mas10,har11,masanew}.
%However, all derivations up to now contained a stubborn piece of mathematical abstractness that could not be reduced to purely information-theoretic terms.   

Recently, a new derivation of Quantum Theory from purely
information-theoretic principles has been presented in Ref.
\cite{qmfrompuri} (see also \cite{game} for a short introduction to
the background).   
In this work, which marks a first step towards the
realization of Wheeler's dream, Quantum Information is shown to
maintain its promise for the understanding of fundamental physics: indeed, the key principle that identifies Quantum Theory is the \emph{Purification Principle} \cite{purification}, which is  directly inspired by the research in Quantum Information.  Quantum Theory is now captured by a complete set of
information-theoretic principles, which can be stated using
\emph{only} the elementary language of systems, processes, and
probabilities.  With respect to related reconstructive works, the new
derivation of Ref. \cite{qmfrompuri} has the advantage of offering a
clear-cut picture that nails down in few simple words what is special
about of Quantum Theory: Quantum Theory is, in the first place, a
theory of information, which shares some basic features with classical
information theory, but differs from it on a crucial point, the \emph{purity
  and reversibility of information processing}. In a standard
set of theories of information, Quantum Theory appears to be the only
theory where the limited knowledge about the processes that we observe in nature
is enough to reconstruct a picture of the physical world where
all processes are pure and reversible.   

More precisely, when we state that Quantum Theory is a theory of information, we mean that the mathematical framework of the theory can be 
expressed by using  only concepts and statements that have an informational significance, such as the concept of signalling, of distinguishability of states, or of encoding/decoding.    Here we refer to ``information" and ``informational significance" in a very basic, primitive sense: in this paper we will not  rely on specific measures of  information, such as the Shannon, Von Neumann, or Renyi entropies.  In fact, the very possibility of defining such quantitative measures is based on the specific mathematical structure of classical and quantum theory (chiefly, on the fact that in these theories every mixed state is a probabilistic mixture of perfectly distinguishable states),   which, for the quantum case, is exactly what we want to pin down with our principles.  

The informational concepts used in this paper are connected to the more traditional language of physics by viewing  the possible physical processes as information processing
events. For example, a scattering process can be viewed as an
event---the interaction---that transforms the input information encoded in the momenta of the incoming particles into the output information encoded in the momenta of the scattered particles. 
   From this perspective, the properties of the particular theory
of information that we adopt immediately translate into properties of our
physical description of the world.  The natural question that we address here is: which properties of a theory of information imply that the description of the world must be quantum? 

The purpose of this paper is to give a short, non-technical answer to the question, providing an account of
the informational principles of Quantum Theory presented in Ref. \cite{qmfrompuri} and of the worldview emerging from them.  Hence, we will focus on the broad picture and
on the connection of the principles with other fundamental areas of
theoretical physics, while referring the reader to the comprehensive
work of Ref.\cite{qmfrompuri} for the mathematical definitions and for
the rigorous proofs of the claims.
 
 \section{A complete set of information-theoretic principles for Quantum Theory}
   
\begin{figure}[h]
  \begin{center}
  \includegraphics[width=.4\textwidth]{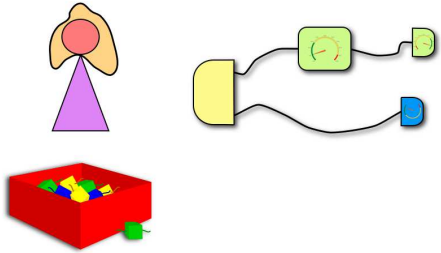}
\caption{ {\bf Alice's laboratory.}   Alice has at disposal many devices, each of them having an input system and an output system (represented by different wires) and possibly a set of outcomes labelling different processes that can take place. The devices can be connected in series and in parallel to form circuits.  A circuit with no input and no output wires represents an experiment starting from the preparation of a state with a given source and ending with some measurement(s).    Specifying a theory for Alice's laboratory means specifying which are the allowed devices and specifying a rule to predict the probability of outcomes in such experiments.} \label{alice}
\end{center}
\end{figure}

To portray Quantum Theory we set up a scene where an experimenter, Alice, has many devices in her laboratory and can connect them in series and in parallel to build up circuits (Fig. \ref{alice}).    In Alice's laboratory, any device can have an input and an output  system, and possibly some outcomes  that Alice can read out.  Each outcome labels a different \emph{process}  transforming the input into the output: the device itself can be viewed as implementing a \emph{random process}.    Some devices have no input: they are \emph{preparations}, which initialize the system in some state.  Other devices have no output: they are \emph{measurements}, which absorb the system and produce an outcome with some probability.

From a slightly more formal point of view, Alice's circuits can be described with a graphical language where boxes represent different devices and wires represent physical systems travelling from one device to the next \cite{purification}, in a way that is inspired by  the picturalist framework by Coecke  \cite{bob}.
These circuits are essentially the same circuits that are commonly used in Quantum Information \cite{nielsenchuang}, except for the fact that here we do not specify from the beginning the mathematical representation of the devices: we do not specify that the possible states are described by density matrices on some complex Hilbert space, or that the possible reversible evolutions are described by unitary operators.   Retrieving these specific mathematical prescriptions from operationally meaningful assumptions   is indeed the main technical point of Ref. \cite{qmfrompuri} and of the other quantum reconstructions  \cite{Har01,maurolast,philip, DakBru09,Mas10,har11,masanew}. 

Since  the devices in Alice's laboratory can have different outcomes, there are two natural ways to associate circuits to an experiment.   
First, a circuit  can represent the schematic of Alice's experimental setup.  For example, the circuit
\begin{equation}\label{withoutoutcomes}
\begin{aligned}
  \Qcircuit @C=1em @R=.7em @! R {  \prepareC {\{\rho_i\}_{i \in  X}} & \qw \poloFantasmaCn {\rm A} & \gate {\{\mathcal C_j\}_{j \in {\rm Y}}} & \qw \poloFantasmaCn {\rm B} &\measureD {\{b_k\}_{k \in Z}}}
\end{aligned}
\end{equation} 
represents a setup where Alice connects a preparation device that outputs system $\rm A$, a transformation device that turns system $\rm A$ into system $\rm B$, and, finally a measurement device that measures system $\rm B$.   Here all the devices are allowed to have outcomes:  outcome $i\in X$ will herald the fact that the first device prepared the state $\rho_i$, $j\in  Y$ will herald that the second device performed the transformation $\mathcal C_j$, and outcome $k\in  Z$ will herald the event $b_k$ in the final measurement.   
In the specific case of Quantum Theory,  $\{\rho_i\}_{i\in  X}$ is going to be an \emph{ensemble of quantum states} of system $\rm A$ (that is, a collection of unnormalized density matrices on a suitable Hilbert space $\mathcal H_{\rm A}$ with the property  $\sum_{i  \in  X }  {\rm Tr} [\rho_i] =1 $),   $\{\mathcal C_j\}_{j \in  Y}$  is  going to be a \emph{quantum instrument} (a collection of completely positive maps sending states on $\mathcal H_{\rm A}$ to states on $\mathcal H_{\rm B}$ with the property that the map $\sum_{j\in Y }  \mathcal C_j$ is trace-preserving), and $\{ b_k\}_{k\in  Z}$ is going to be a \emph{quantum measurement} (a collection of positive operators on $\mathcal H_B$ with the property $\sum_{k\in Z}   b_k  =  I_{\rm B}$, the identity on $\mathcal H_{\rm B}$).   A reader who is not familiar with these notions can find a didactical presentation in chapter 8 of Ref. \cite{nielsenchuang}.  Note that the graphical representation of the circuit has a privileged direction (from left to right in our convention), this direction corresponding to the \emph{input-output arrow}:  wires on the left of a box represent its inputs, wires on the right of a box represent its  outputs.  Such a prefereed input-output arrow will be important later in the statement of the Causality principle. 

The second way to associate a circuit to an experiment is to represent  the instance of the experiment corresponding to a particular sequence of outcomes. For example, the circuit
\begin{equation}\label{withoutcomes}
\begin{aligned}
  \Qcircuit @C=1em @R=.7em @! R {  \prepareC {\rho_i} & \qw \poloFantasmaCn {\rm A} & \gate {\mathcal C_j} & \qw \poloFantasmaCn {\rm B} &\measureD {b_k}}
\end{aligned}
\end{equation} 
represents a particular instance of the experiment with the setup in Eq. (\ref{withoutoutcomes}), corresponding to the particular sequence of outcomes $(i,j,k)$.  In this specific instance, the first device has prepared  the state $\rho_i$, the second device has implemented the transformation $\mathcal C_j$, and the final measurement has given outcome $z$.  A circuit with no open wires, like the circuit in Eq. (\ref{withoutcomes}), will be associated to a joint probability $p( \rho_i,  \mathcal C_j,  b_k )$, namely the joint probability of obtaining the outcomes $(i,j,k)$ in the experiment with setup (\ref{withoutoutcomes}).   Notice however that nothing prevents us from drawing circuits with open wires, such as  
\begin{equation}\label{nondemolition}
\begin{aligned}
  \Qcircuit @C=1em @R=.7em @! R {  \prepareC{\rho} & \qw \poloFantasmaCn {\rm A} & \multigate{1} {\mathcal U} & \qw \poloFantasmaCn {\rm A} &\qw  \\
   \prepareC{\sigma} & \qw \poloFantasmaCn {\rm P} & \ghost{\mathcal U} & \qw \poloFantasmaCn {\rm P} &\measureD{m_i}},
\end{aligned}
\end{equation} 
which represents a ``non-demolition measurement", where the system $\rm A$ (initially in the state $\rho$) interacts with a probe $\rm P$ (initially in state $\sigma$) through some transformation $\mathcal U$, after which the probe undergoes a measurement, giving outcome $i$. 

In summary, our basic framework to treat general theories of information is based on the combination of the graphical language of circuits with elementary probability theory. Such a combination of circuits and probabilities, originally introduced in Ref. \cite{purification} and discussed in Ref. \cite{hardyfoil},  offers a simple ground for the study of generalized probabilistic theories \cite{pr,Har01,barrett,nobroad,maurolast,wilce}, and allows one to avoid some of the technicalities of the more traditional  ``convex sets framework", such as the choice to the tensor product (see e.g. \cite{wilce}).   

The features of the probability distributions arising in Alice's experiments  depend on the particular physical theory describing her laboratory:  At this basic level, the theory could be classical or quantum, or any other fictional theory that we may be able to invent.  We now start restricting the circle of possible theories:  first of all, we make sure that Alice's laboratory is not in a fictional Wonderland, but in a standard world enjoying some elementary
%by requiring some elementary 
properties common to Classical and Quantum Theory.   The first property is:  

\begin{prin}[Causality]  
The probability of an outcome at a certain step does not depend on the choice of experiments performed at later steps. 
\end{prin}

The word {\em later} in the statement of the principle refers to the  ordering of the computational steps in a circuit induced by the input-output connections:   in our
graphical representation the ordering goes from the left to the right and a box connected to the output of another represents a later computational step [cf. Eqs. (\ref{withoutoutcomes}) and (\ref{withoutcomes})]. The causality principle identifies the input-output ordering of a circuit with the {\em causal
  ordering}, namely the direction along which information flows,
without any refluence. In more physical terms, we could informally replace the word ``step'' with the word ``time'' in the
formulation of causality.   In this language, Causality is the requirement that Alice's future choices do not affect the outcomes of her present experiments \emph{(no-signalling from the future)}.

Causality is implicit in the framework in most works in the tradition of generalized probabilistic theories \cite{pr,Har01,barrett,nobroad,DakBru09,Mas10,wilce}. 
The reason why we are stating it explicitly as the first principle of our list is that we would like it to be
a reminder that the formulation of Quantum Theory, in the way it is presently known,
requires a well-defined causal structure in the background. This immediately opens the question
whether it is possible to formulate a general version of Quantum
Theory in scenarios where such a well-defined causal structure cannot be
taken for granted.  As it was observed by Hardy
\cite{causaloid}, the formulation of such a generalized Quantum Theory
with indefinite causal structure could be a route to the formulation
of a quantum theory of gravity.  In this spirit, the
information-theoretic principles presented here are very appealing,
because they suggest to construct a generalized Quantum Theory on indefinite causal structure by
weakening the Causality principle while keeping the other principles
unaltered.

Let us set more requirements on the processes taking place in Alice's laboratory. 
 For every random process, there is also a \emph{coarse-grained process} where some random outcomes are joined together, thus neglecting some information.  A \emph{fine-grained process} is instead a process where no information has been neglected: in this case Alice has maximal knowledge about the process  taking place in her laboratory.    For example, in the roll of a die  the fine-grained processes are ``the roll yielded the number $n$", with $n=1,2,3,4,5,6$, while ``the roll yielded an  even number" is a coarse-grained process:  When Alice declares outcome ``even" she is joining together the outcomes 2, 4, and 6, thus neglecting the corresponding information.  For preparation processes,  the coarse-grained processes are called \emph{mixed states} and fine-grained processes are called \emph{pure states}.

 Our second principle is:  
\begin{prin}[Fine-Grained Composition] 
The sequence of two fine-grained processes is a fine-grained process.  
\end{prin}
This principle establishes  that \emph{``maximal knowledge of the episodes implies maximal knowledge of the history"}: if  Alice possesses maximal knowledge about all processes in a sequence, then she also possesses maximal information about the whole sequence.  A physical theory where this did not hold would be highly pathological, because the mere composition of two processes, which considered by themselves are specified with the maximum degree of accuracy possible,  would generate some global information that cannot be accessed on a step-by-step basis.  For preparation processes, this would mean that by putting together two systems that individually are in a pure state, we would get a compound system  that, considered as a whole, is in a mixed state.  We will come back to this point in more detail in the discussion of our fifth principle, Local Tomography, which has a similar, but yet different and logically independent content.

If Alice describes the system as being in a pure state, then this means that she has maximal knowledge about the system's preparation.  Instead, if Alice describes the system as being in  a mixed state, then she is ignoring (or choosing to ignore) some information about the preparation. When  Alice describes the preparation of her system with a mixed state $\rho$,  her description is compatible with the system being prepared in any of the  pure states from which $\rho$ results as a coarse-graining.     This concept can be easily exemplified for the roll of a (generally unfair) die: here the pure states are  numbers from $1$ to $6$, while the mixed states  are probability distributions over $\{1, \dots, 6\}$.   A mixed state $p$ is compatible with every pure state $x\in  \{1,\dots, 6\}$ such that $p(x)  >  0$, while it is not compatible with those  $x$  such that $p(x)  = 0$.   If a mixed state $p$ is not compatible with some pure states $x \in  X_0$, then     
  it is possible to distinguish perfectly between $p$ and any other probability distribution  $q$ that has support contained in $X_0$.
 The same feature holds in Quantum Theory:  if a density matrix $\rho$  on some Hilbert space $\mathcal H$  is not compatible with some pure state $\varphi$ [that is, if there is no probability $p>0$ and no density matrix $\sigma$ such that $\rho  =   p  |\varphi\rangle\langle \varphi|  +  (1-p)  \sigma$]  then the the density matrix $\rho$ should have a non-trivial \emph{kernel}, defined as the set of  all vectors $|\psi\rangle  \in  \mathcal H$  such that $\langle \psi|\rho  |\psi\rangle  = 0$.   Hence $\rho$,  will be perfectly distinguishable from any pure state $|\psi\rangle$  in its kernel, and, more generally, from any mixture of pure states in its kernel.  Abstracting from these specific examples, we can state the following general principle:
 
%Some states are so mixed that, according to them, the system could be in any pure state. %:  in this case Alice is completely ignorant about the preparation.  
%We  call such states \emph{completely compatible} and stipulate the following:    

\begin{prin}[Perfect Distinguishability]  
If a state is not compatible with some preparation, then it is perfectly distinguishable from some other state.
\end{prin}

In other words, % \emph{``if Alice's information not compatible with some preparation, then %specify the value of a bit"}.
%there is a yes-no property that she can predict  a with certainty"}:
%  she can use it to communicate the value of a bit to another experimenter Bob"}.  
%Indeed, our principle guarantees that Alice's state, call it $\rho_0$, is perfectly distinguishable from some other state, call it $\rho_1$:  the proposition "the system is in state $\rho_1$"  can be experimentally falsified in a one shot experiment. 
\emph{``possessing definite information about the preparation implies the ability to experimentally falsify some proposition"}.    
Indeed, suppose that knowing that the system is prepared in the state $\rho_0$ allows us to exclude that the system is in a pure state $\varphi$. Then, Perfect Distinguishability guarantees that $\rho_0$ is perfectly distinguishable from some other state, call it $\rho_1$.  The proposition ``the system was prepared in the state $\rho_1$" can then be falsified by performing the measurement that distinguishes perfectly between $\rho_0$ and $\rho_1$.  
Note that, thanks for Perfect Distinguishability,  Alice can use $\rho_0$ and $\rho_1$ to encode  the value of a classical bit in a physical support without errors.

\begin{figure}[h]
 \begin{center} \includegraphics[width=.4\textwidth]{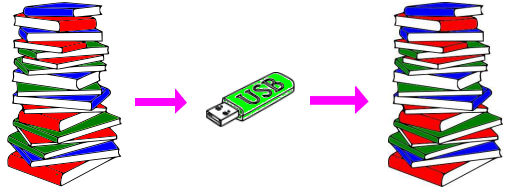}
 \end{center}
\caption{{\bf Compressing information.}   Alice  encodes information (here represented by a pile of books) in a suitable system  carrying the smallest possible amount of data (here a USB stick).    The most advantageous situation is when the compression is \emph{lossless} (after the encoding Bob is able to perfectly retrieve the information) and \emph{maximally efficient} (the encoding system contains only the pure states needed to convey the information compatible with $\rho$).     } \label{compression}
\end{figure}

Suppose that Alice wants to transfer to another experimenter Bob all the information she possesses about a system.   If the system's state $\rho$ is   mixed, then Alice ignores the exact preparation: with some non-zero probability the system could be in any of the pure states compatible with $\rho$. Hence,  in order for her transmission to be successful, the transmission should work for every pure state compatible with $\rho$.  
  Moreover, since transferring data has a cost, Alice would better \emph{compress the information} (Fig. \ref{compression}).    

Our fourth principle guarantees the possibility of such an ideal compression: 
\begin{prin}[Ideal Compression]  
Information can be compressed in a lossless and maximally efficient fashion. 
\end{prin}
Due to the Ideal Compression principle, Alice can transfer information without transferring the particular physical system in which information is embodied.  In the example of the roll of the die, Ideal Compression principle can be illustrated as follows:  if our information about the outcome of the roll is described by a probability distribution $p$ with $p(1)  =  p(2)  =\frac 12$ and  $p(3)  =  p(4)  =p(5) = p(6) =0$, then we can faithfully encode this information in the state of a coin, by encoding $1$ into ``heads" and $2$ into ``tails".  This compression is perfectly lossless and maximally efficient in the sense of our definition.   Note that this elementary notion of ideal compression differs from the more articulate notion used in Shannon's theory \cite{shannon}, in Schumacher's quantum theory of compression, and in everyday information technology, where one is often willing to tolerate some losses in order to further reduce the size of the physical support in which information is encoded. In that case, the compression is required to be lossless only in the asymptotic limit of many identical uses of the same information source, and the efficiency is defined among the set of  compression protocols that are asymptotically lossless  \cite{shannon,schumaker}.

The next principle concludes  our list of requirements that are satisfied both by Classical and Quantum Theory:  
\begin{prin}[Local tomography]  
The state of a composite system is determined by the statistics of local measurements on the components.
\end{prin}
\begin{figure}[h]
  \begin{center}\includegraphics[width=.4\textwidth]{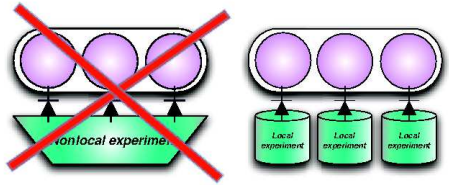}
\end{center}
\caption{ {\bf Local Tomography.}  Alice can reconstruct the state of  compound systems using only local measurements on the components.  A world where this property did not hold would contain global information that cannot be accessed with local experiments.  } 
\end{figure}

Local Tomography plays a crucial role in reducing the complexity of experimental setups needed to characterize the state of multipartite systems, ensuring that there all the  information contained in a composite system is accessible to joint local measurements.  Mathematically, this principle is the key reason for the choice of \emph{complex}  (instead of real) Hilbert spaces: in real Hilbert space Quantum Theory there are some bipartite states that can be distinguished perfectly with global measurements, but give the same statistics for all possible local measurements, as it was noted by Wootters \cite{woottersloc}.    It is worth noticing that Quantum Theory on real Hilbert spaces still satisfies the Local Tomography principle \emph{if we restrict our attention to pure states} \cite{purification}.   Finally, it is  interesting to comment  on the relation between Fine-Grained Composition and Local Tomography. Although these two principles have a similar flavour (both of them exclude the possibility of having some inaccessible global information), they are actually very different.  Fine-Grained Composition states that if we put together two processes of which we have maximal knowledge, then we obtain a process of which we have maximal knowledge as well.     In particular, for preparation processes this means that if we prepare two systems $\rm A$ and $\rm B$ in two pure states, then the composite system $\rm A B$ will be in a pure state as well.   This is a much weaker statement than Local Tomography! Indeed, it is quite simple to see that Quantum Theory on real Hilbert spaces satisfies Fine-Grained Composition, but not Local Tomography.    In principle, it is also conceivable to have fictional theories that satisfy Local Tomography, but not Fine-Grained Composition: although Local Tomography implies Fine-Grained composition in the particular case of preparation processes, it is possible to construct locally tomographic theories where Fine-Grained Composition fails at the level of general processes (processes that have both a non-trivial input and a non-trivial output).

The five principles presented so far define a family of theories of information that can be regarded as a standard. If it were just for these principles, Alice's experiments could still be described, for example, by Classical Theory.   
What is then special about Quantum Theory? What makes it different from any other theory of information satisfying the five basic principles presented so far?   Our answer is the following:  Quantum Theory is the only theory of information that is compatible with a description of  physical processes only in terms of pure states and reversible interactions.      In a sense, Quantum Theory is \emph{the only physical theory of information}: the only theory where Alice's ignorance  about processes happening in her laboratory is compatible with a complete picture of the physical world.  Colourfully reinterpreting Einstein's quote: God does not play dice, but we definitely do, and God must be able to describe our game!
   
Let us spell out our last principle precisely.   In Quantum Theory, every random process can be simulated as a reversible interaction of the system with a pure environment (i.e. with an environment in a pure state). This simulation is \emph{essentially unique}:  once we fix the environment,  two simulations of the same random process can only differ by a reversible transformation acting on the environment.  Essential uniqueness is a very important feature:  it means that Alice's information about a random process happening in her laboratory is sufficient for her to determine the system-environment interaction in the most precise way possible (compatibly with the fact that Alice has no access to the environment).  Distilling these ideas in a principle, we obtain the following:

%Let us spell out our last principle precisely.   In Quantum Theory, every random process can be simulated as a reversible interaction of the system with an environment in a pure state. This simulation is \emph{essentially unique}:  once we fix the environment,  two simulations of the same random process can only differ by a reversible transformation acting on the environment.  This means that Alice's information about a random process
% is sufficient to determine the system-environment interaction in the most precise way possible (compatibly with the fact that she has no access to the environment).  Distilling these ideas in a principle: %, we obtain % the following:  

\begin{prin}{\bf (Purity and Reversibility of Physical Processes)} \emph{Every random process can be simulated  in an essentially unique way as a reversible interaction of the system with a pure environment.} 
\end{prin}
The Purity and Reversibility principle is closely connected with the idea of \emph{reversible computation}, introduced in the seminal works by Bennett \cite{revbennet} and Fredkin-Toffoli \cite{fredtof}.   In the world of classical computers, it was shown that  every deterministic function (even a non-invertible function) can be computed in a reversible way, by suitably enlarging the space of the computation with additional bits initialized in a fixed pure state.    This is a fundamental observation because it hints at the possibility of computing without erasing information, which, by Landauer's principle \cite{landauer}, would imply an energy cost and an increase of entropy in the environment [see also pp. 153-161 of \cite{nielsenchuang} for an easy introduction to these topics].   In the classical world, however, only deterministic functions can be computed through a reversible interaction of the input system with a pure environment, whereas classical stochastic processes require the  environment to be initialized in a mixed state.  In other words, the realization of classical stochastic processes requires a source of randomness in the environment, which, loosely speaking, has to ``pump entropy" into the system).  This is unfortunate, because stochastic processes are also computationally interesting and useful for a number of applications in the the most disparate disciplines (think e.g. of the wide application of the Montecarlo and Metropolis algorithms).   Instead, the bonus offered by Quantum Theory, as stated by the Purity and Reversibility principle, is that \emph{every allowed process} (including those of a stochastic nature) can be realized in a pure and reversible fashion, thus allowing for a fully reversible model of information processing.    

The Purity and Reversibility principle concludes our list.  For finite systems (systems whose state is determined by a finite number of outcome probabilities)  the six principles  presented above describe Quantum Theory completely \cite{qmfrompuri}: complex Hilbert spaces,  superposition principle, Heisenberg's uncertainty relations,  entanglement,  no-cloning,  teleportation, violation of Bell's inequalities, quantum cryptography---every quantum feature is already here, encapsulated in the principles. The detailed proof can be found in Ref. \cite{qmfrompuri}.   
%Moreover, 
The surprising result here is that, although our sketch of Alice's laboratory may seem  too simplistic, especially to physicists (after all, the Universe is not a big laboratory where we can choose the preparations and measurements at will!),  this scenario is rich enough to capture the basic language of Quantum Theory.   Technically, our information-theoretic principles imply the following mathematical statements:  
\begin{itemize}
\item physical systems are associated to complex Hilbert spaces
\item  the maximum number of perfectly distinguishable states of the system is equal to the dimension of the corresponding Hilbert space
\item  the pure states of a system are described by the unit vectors in the corresponding Hilbert space (up to a global phase)
\item the reversible processes on a system are described by the unitary operators on the corresponding Hilbert space (up to a global phase)
\item the measurements on a system are described by resolutions of the identity in terms of positive operators  $\{P_i\}_{i  \in X}$  on the corresponding Hilbert space  (aka POVMs \cite{nielsenchuang})
\item the mixed states of a system are described by density matrices  on the corresponding Hilbert space 
\item the probabilities of outcomes in a measurement are given by the Born rule  $p_i  =   {\rm Tr} [P_i \rho   ]$, where $\rho$ is the density matrix representing the system's state and $\rm Tr$ denotes the trace of a matrix
\item the Hilbert space associated to a composite system is the tensor product of the Hilbert spaces associated to the components.
\item random processes are described by completely positive trace-preserving maps
\end{itemize}
Remarkably, these  statements are exactly the mathematical features mentioned in the original paper by Fuchs \cite{fuchs}, which was calling for an information-theoretic reason thereof.  % in the an eloquent table entitled ``Quantum Mechanics: The Axioms and our Imperative!".  

Although the derivation of Ref. \cite{qmfrompuri} holds for finite systems, it is natural to expect that the principles discussed here will identify Quantum Theory also in infinite dimension:  in that case one has to take care of many technicalities,  which however have more to do with the mathematical problem of infinity rather than with the conceptual problems of  Quantum Theory.

% Surprisingly, although our sketch of Alice's laboratory may seem  too simplistic  (after all, the Universe is not a big laboratory!),  this scenario is rich enough to capture Quantum Theory. 

%{\bf more about purity and reversibility} 

\section{Conservation of Information and the Purification Principle}

We now illustrate two important messages of the Purity and Reversibility Principle.  The first message is that irreversibility can be always modelled as  loss of control over an environment.  In other words, the principle states a law of \emph{Conservation of Information} according to which information can never be destroyed but can only be discarded.  Here we are talking about information in a basic, non-quantitative sense: we mean information about the system's preparation, which is encoded in the system's state and allows one to predict the probabilities of outcomes in all the experiments one can perform on the system.    Consistently with this definition, we say that the information encoded in the system's  state is conserved by a process if and only if after the process the system can be taken back to its initial state.  If we regard the pieces of information carried by physical systems as  fundamental blocks constituting our world, then  the Conservation of Information is a  must.   
  Its importance, at least at the heuristic level, can be easily seen in the debate that followed Hawking's discovery of the thermal radiation emitted by black holes \cite{hawk}: The trouble with Hawking's result was exactly that it seemed to negate the Conservation of Information \cite{preskill}.  In this case, the conviction that the Conservation of Information is fundamental led t'Hooft \cite{holotofft} and Susskind \cite{holosuss} to the formulation of the holographic principle, a major breakthrough in quantum gravity and quantum field theory.

The second important message of the Purity and Reversibility Principle is that we can simulate every physical process using a \emph{pure} environment, that is, without pumping entropy from the environment. Again, here we are talking about entropy in a very basic sense:  whichever quantitative definition we may choose, entropy  must be zero for pure states  and non-zero for mixed states.     We already discussed the significance of the purity requirement for reversible computation, in the spirit of the works by Bennett \cite{revbennet},  Fredkin and Toffoli \cite{fredtof} and in connection with Landauer's principle \cite{landauer}.

%{\bf purification, schroedinger}  
Purity and Reversibility can be expressed in an elegant way as  \emph{Purification Principle}: \emph{``every mixed state arises in an essentially unique way by discarding one component of a compound system in a pure state"} \cite{purification}.    The Purification Principle is the statement that  the ignorance about a part is always compatible with the maximal knowledge about the whole, a statement that is very closely connected with the ideas of Schr\"odinger about entanglement (cf.  the statement  ``another way of expressing the peculiar situation is: the best possible knowledge of a \emph{whole} does not necessarily include the best possible knowledge of all its \emph{parts}" in Ref. \cite{Schr35}).         
Using this language, our result can be rephrased  as:  \emph{quantum theory is the unique theory of information where the ignorance about a part is compatible with the maximal knowledge about the whole}.   This result finally realizes and \emph{proves}  in a mathematically precise way the intuition expressed by Schr\"odinger with his prophetic words about entanglement: ``I would not call that \emph{one} but rather \emph{the} characteristic trait of quantum mechanics, the one that enforces its entire departure from classical lines of thought" \cite{Schr35}. 

Remarkably, the compatibility of the ignorance about a part with the maximal knowledge about the whole is also the key idea in a recent proposal for the foundations of statistical mechanics \cite{statmec}, where it has been shown that  the state of a small subsystem of a composite system in a random \emph{pure} state will be described by the microcanonical ensemble (i.e. by the maximally mixed state)  with high probability.   In addition to this and to the already mentioned relation with reversible computation, it is worth noting that the Purification Principle has countless applications in Quantum Information, ranging from the security analysis of quantum cryptographic protocols to the study of coding schemes in quantum Shannon theory, from the definition of distinguishability measures such as the fidelity and the diamond norm to the theory of quantum error correction  (we refer the reader to the Refs. \cite{nielsenchuang,preskillnotes,watrous,wilde} for a didactical presentation of many of these topics).  The purification principle has also direct applications in quantum  estimation and quantum metrology \cite{refframe,giuliolec,davidovich}.

\section{Discussion and conclusions}

Before concluding, some remarks are in order. First of all, it is important to stress that the principles in Ref. \cite{qmfrompuri} are about the syntax of physical experiments, and not about  their semantics.  When we discuss about systems, transformations, and measurements, we take a general information-theoretic standpoint that abstracts from the specific physical realization of these notions. From the information-theoretic standpoint,  all two-level systems are equivalent, no matter if they are implemented with the polarization of a photon, the magnetic moment of a nucleus, or the charge  in a superconductor.  This is at the same time a strength and a limitation of the information-theoretic approach.  On the one hand,   forgetting about the specific details of the physical implementation is a very powerful abstraction: it is the abstraction that allows us to talk about ``software" without  specifying the details of the ``hardware", and to prove high-level statements that are implementation-independent (think, for example to the no-cloning theorem \cite{wootterszurek,dieks}).  On the other hand, in physics it is also fruitful to attach a specific physical meaning to the abstract information-theoretic entities of the theory:  for example, among all possible measurements, one would like to single out a particular one as the measurement of  the ``energy" or another one as the measurement of ``angular momentum".  Likewise, among all allowed states of the system, one would like to know which ones are ``ground states of the energy", or which ones are states where  ``the angular momentum is aligned in the $x$ direction".  
The basic information-theoretic framework of Ref. \cite{qmfrompuri} does not address these issues: to include physical notions like  ``energy", ``angular momentum", ``polarization", ``mass", ``charge", ``position", ``velocity", one would have to enrich to the basic language in which our principles are phrased.  There is no doubt that this is a very worthwhile thing to do, because, all in all,  physical laws are  quantitative relations involving these notions.   
However, one important lesson of Ref. \cite{qmfrompuri} (and, more generally of the recent information-based quantum reconstructions  \cite{DakBru09,Mas10,har11}) is that the basic mathematical structure of Quantum Theory can be completely characterized without referring to traditional physical notions such as ``position", ``velocity", or ``mass".    

The difference between the information-theoretic syntax and  physical semantics can be well exemplified by discussing how much of the Schr\"odinger equation can be reconstructed in the information-theoretic approach. As we already mentioned, from our principles we can derive that  the reversible transformations of a system are described by unitary operators on the corresponding Hilbert space.  As a consequence,  a reversible time-evolution in continuous time  will be described by a family of unitary transformations $U_t  ,   t  \in  \mathbb R$.   It is then immediate to show that the unitaries should satisfy the equation $ i \frac {  {\rm d  } }{{\rm d} t}  U_t  =  H(t)  U_t$, where $H(t)$ is some Hermitian operator that we can call the ``Hamiltonian" of the system.  This is exactly the mathematical structure of the Schr\"odinger's equation.  However, the physical interpretation of $H$ as the ``energy" of the system is not included in the information-theoretic framework, but instead it is part of the physical content of the Schr\"odinger equation.  Likewise,  it is important to note that  in our framework there is no fundamental scale: no ``far vs.~ close",  nor ``slow vs.~fast".  Again,  the actual value of the Plank's constant $\hbar$ is part of the physical semantics of Quantum Mechanics, and not of the basic syntax of Quantum Theory.  

It is important to note that also the very scope of the information-theoretic derivations focuses on the syntax, rather than on the semantics:   Questions like ``What is an observer?" or  ``What is a measurement?"  are not addressed by the principles.   Neither Ref. \cite{qmfrompuri} nor the other reconstruction works \cite{DakBru09,Mas10,har11,masanew}  aim to solve the measurement problem or any related interpretational issue.

%{\bf the big picture, again}  

In conclusion,  building on the results of Ref. \cite{qmfrompuri}, in this paper we presented  six informational principles that completely capture the world of Quantum Theory. 
The theory can now be described with the elementary language of Physics,   without appealing to external \emph{ad hoc} notions. 
The view emerging from the principles is that Quantum Theory is %, in the first place,  \emph{ a theory of information}. 
%---stating that information propagates forward in time,   that maximal information about the episodes implies maximal information about the history, that non-completely mixed states allow to perfectly communicate the value of a bit, that information can be compressed, and that global information can be retrieved with local measurements.    
%Specifically, Quantum Theory  is 
  \emph{the only physical theory of information}: the only theory where the limited information possessed by  the experimenter %has about the observed processes
   is enough to construct a picture of the world where  all states are pure and all processes are reversible.

Now that our portrait of Quantum Theory has been completed, a natural avenue of future research consists in exploring the alternative theories that are allowed if we relax some of the principles.  Given the structure of   our work, which highlights  Purity and Reversibility as ``the characteristic trait" of Quantum Theory,  it becomes interesting to study theories in which one weakens some of the first five (standard) principles while keeping Purity and Reversibility.  All these alternative theories could be rightfully called ``quantum", for they share with the standard Quantum Theory its distinctive feature.  One natural weakening of the principles would be to relax Local Tomography, thus allowing  Quantum Theory on  real Hilbert spaces, an interesting toy theory which exhibits quite peculiar information-theoretic features \cite{wootterssharing}.   More challenging and more exciting at the same time would be to venture in the realm of non-causal theories that satisfy the Purity and Reversibility principle, a much broader family of theories that are interesting in view of a formulation of quantum theory in the absence of a definite causal structure.  The study of quantum theories with indefinite causal structure is a completely new avenue of research that has just begun to be investigated  \cite{hardycomp,switch,brukner,giulio,programmable}, and we believe that it will lead to the discovery of new quantum effects and interesting information processing protocols.   
       
\medskip 
%GC is grateful to B Coecke for stimulating discussions and to A Hamma for suggestions on an earlier version of this paper.  
%{\bf Acknowledgments.}  
%\medskip
%{\bf Author contributions.}  All authors contributed to the initial conception of the ideas and to the writing of the manuscript.

 \section*{Acknowledgements}
GC acknowledges support from  the National Basic Research Program of China (973)
2011CBA00300 (2011CBA00301) and from Perimeter Institute for Theoretical Physics  in the initial stage of this work.   Research at QUIT has been supported by the EC through the project COQUIT.     Research at Perimeter Institute for Theoretical Physics is supported in part by the Government of
Canada through NSERC and by the Province of Ontario through MRI.  We acknowledge the three anonymous Referees of this paper for valuable comments that have been useful in improving the original manuscript.
%==========================================================
%==========================================================
% Back Matter (References and Notes)
%----------------------------------------------------------
% Style and layout of the references
\bibliographystyle{mdpi}
\makeatletter
\renewcommand\@biblabel[1]{#1. }
\makeatother
%----------------------------------------------------------
% Use the following option to include external BibTeX files:
%\bibliography{template}

\begin{thebibliography}{55}
\bibitem{qmfrompuri}  Chiribella, G.,  D'Ariano, G. M., and Perinotti, P. Informational Derivation of Quantum Theory,    {\em Phys. Rev. A} {\bf 2011}, {\em 84}, 012311.  
\bibitem{wheelerNYT}  Wheeler,  `A Practical Tool', but Puzzling Too,  {\em New York Times}, {\bf 2000}, December 12. 
\bibitem{redei} Redei, M.    Why John von Neumann did not like the Hilbert space formalism of quantum mechanics (and what he liked instead),  {\em Stud. Hist. Phil. Mod. Phys.} {\bf 1997}  {\em 27}, 493-510.
\bibitem{von32} von Neumann, J.   {\em Mathematical Foundations of
    Quantum Mechanics}; Princeton University Press: Princeton, 1932.
\bibitem{BirkVN36} Birkhoff, G. and von Neumann, J.,  The Logics of Quantum Mechanics, {\em Ann. Math.}  {\bf 1936} {\em  37}, 823-843.
\bibitem{wheelerIt} Wheeler, J. A.   Information, Physics, Quantum: The Search for Links.  In \emph{Complexity, Entropy, and the Physics of Information};  Zurek, W. Ed.; Addison-Wesley: Redwood City, CA, 1990; p. 5.
\bibitem{woot}  Wootters, W. K.,  The Acquisition of Information from
Quantum Measurements, PhD thesis, University of Texas
at Austin, 1980.
\bibitem{weiz}  von Weizsacker, C. F. The Structure of Physics,   G\"ornitz, T. and Lyre, H. Eds.: Springer:  Dodrecht, 2006.   
\bibitem{zei} Zeilinger, A. A Foundational Principle for Quantum Mechanics,  {\em Found. Phys.}  {\bf 1999}  {\em 29}, 631-643.
\bibitem{bruzei}  Brukner, \v C, and Zeilinger, A.  Information and fundamental elements of the structure of quantum theory, 	in \emph{Time, Quantum, Information}, Castell, L.  and Ischebeck, O.,  Springer: Berlin Heidelberg, 2003; pp. 323-354.
\bibitem{wootterszurek}   Wootters, W. K. and  Zurek, W. H.   A Single Quantum Cannot be Cloned, {\em Nature} {\bf   1982}  {\em  299}, 802-803.   
\bibitem{dieks} Dieks, D.  Communication by EPR devices,  {\em Phys. Lett. A}  {\bf   1982},  {\em 92}, 271-272.   
\bibitem{tele} Bennett, C. H., Brassard,  G.,  Cr\'epeau, C.,  Jozsa,  R.,   Peres, A., and  Wootters,  W. K. Teleporting an Unknown Quantum State via Dual Classical and Einstein-Podolsky-Rosen Channels, {\em Phys.
  Rev. Lett.} {\bf (1993)}, {\em  70}, 1895-1899.
\bibitem{bb84}   Bennett, C. H. and  Brassard, G.  Quantum Cryptography: Public key distribution and coin tossing, in {\em Proceedings of the IEEE International Conference on Computers, Systems, and Signal Processing}, Bangalore, India, 1984; pp. 175-179.
 \bibitem{e91}   Ekert, A. K.   Quantum Cryptography Based on Bell's Theorem, {\em Phys. Rev.  Lett.} {\bf 1991},  {\em 67},  661-663. 
 \bibitem{grover}   Grover, L. K. A Fast Quantum Mechanical Algorithm for Database Search, Proceedings of 28th Annual ACM Symposium on the Theory of Computing (STOC), 1996,  pp. 212-219. 
 \bibitem{shor}  Shor, P. W.  Polynomial-Time Algorithms for Prime Factorization and Discrete Logarithms on a Quantum Computer, {\em SIAM J. Comput.}  {\bf 1997},  { \em 26}, 1484-1509.
\bibitem{fuchs} Fuchs, C. A.  Quantum Mechanics as Quantum
    Information, Mostly.  {\em J. Mod. Opt.}  {\bf 2003}, {\em 50}, 987-1023.
\bibitem{brassard}    Brassard, G. Is Information the Key?, {\em Nature Physics} {\bf 2005}, {\em 1}, 2-4.
\bibitem{infocau} Pawlowski, M., Paterek, T., Kaszlikowski, D., 
  Scarani,  V., Winter,  A. and  Zukowski, M.  Information Causality as
    a Physical Principle, {\em Nature} {\bf 2009}, {\em 461}, 1101-1104.
\bibitem{Har01}  Hardy, L. Quantum Theory From Five Reasonable Axioms, arXiv:quant-ph/0101012.   
\bibitem{maurolast}  D'Ariano, G. M.   Probabilistic Theories: What is Special about Quantum Mechanics?, in \emph{Philosophy of Quantum Information and Entanglement},
   Bokulich, A. and Jaeger, G., Eds.; Cambridge University Press: Cambridge UK, 2010; pp. 85-126.
   \bibitem{philip}  Goyal, P. , Knuth, K. H., and  Skilling, J. Origin of Complex Quantum Amplitudes and Feynman's Rules, {\em Phys. Rev. A} {\bf 2010} {\em 81}, 022109.  
  \bibitem{DakBru09} Dakic, B.  and  Bruckner, \v C.  Quantum Theory and Beyond: Is Entanglement Special?,   in {\em Deep Beauty: Understanding the Quantum World through Mathematical Innovation}, Halvorson, H., Ed.;  Cambridge University Press: Cambridge, 2011, pp. 365-392.
\bibitem{Mas10} Masanes, L.  and M\"uller, M., A derivation of quantum theory from physical requirements,  {\em  New J. Phys.} {\bf 2011},  {\em 13}, 063001.
\bibitem{har11} Hardy, L. Reformulating and Reconstructing Quantum Theory, http://arxiv.org/abs/1104.2066 
\bibitem{masanew}  Masanes,  L., Mueller,  M. P., Augusiak,  R., and Perez-Garcia, D., A digital approach to quantum theory, arXiv:1208.0493.
\bibitem{game}   Brukner, \v C.  Questioning the Rules of the Game, {\em Physics} {\bf 2011}, {\em 4}, 55.   
\bibitem{purification}  Chiribella, G.,  D'Ariano,   G. M.,   and Perinotti, P.  Probabilistic Theories with Purification, {\em Phys. Rev. A} {\bf 2010},  {\em 81}, 062348.
\bibitem{bob} Coecke, B.,  Quantum picturalism, {\em Contemporary Physics} {\bf 2010},  {\em 51}, 59-83.
\bibitem{nielsenchuang}  Nielsen, M. A. and Chuang, I. L. {\em Quantum Computation and Quantum Information},   Cambridge University Press: Cambridge, 2000.
\bibitem{hardyfoil}  Hardy, L.  Foliable Operational Structures for General Probabilistic Theories, in  {\em  Deep Beauty: Understanding the Quantum World through Mathematical Innovation}, Halvorson, H. Ed.  Cambridge University Press: Cambridge, 2011, p. 409.  
\bibitem{pr}   Popescu, S. and Rohrlich, D.  Quantum Nonlocality as an Axiom, {\em Found. Phys.}  {\bf 1994}  {\em 3},    379-385.
\bibitem{barrett} Barrett, J.   Information Processing in Generalized Probabilistic Theories, {\em Phys. Rev. A}  {\bf 2007}  {\em 75}, 032304.
\bibitem{nobroad}  Barnum, H., Barrett, J.,  Leifer, M., and  Wilce, A. A Generalized No-Broadcasting Theorem, {\em Phys. Rev. Lett.} {\bf 2007}  {\em  99}, 240501.
\bibitem{wilce} Barnum, H. and Wilce, A, Information Processing in Convex Operational Theories, {\em Electronic Notes in Theoretical Computer Science} {\bf 2011} {270}, p.3-15.  
\bibitem{causaloid}  Hardy, L.     Towards Quantum Gravity: a Framework for Probabilistic Theories with Non-fixed Causal Structure,   {\em J. Phys. A} {\bf 2007} {\em 40}, 3081-3099.  
\bibitem{shannon}   Shannon, C. E.   A Mathematical Theory of Communication, {\em Bell Sys. Tech. Jour.},   {\bf 1949}, {\em 27}, 379Ð423,  623-656.
\bibitem{schumaker}  Schumaker, B.  Quantum coding, {\em Phys. Rev. A}   {\bf 1995} {\em  51}, 2738- 2747.  
\bibitem{woottersloc} Wootters, W. K. Local Accessibility of Quantum States, in {\em Complexity, Entropy and the Physics of Information},  Zurek, W. H. (Ed.), Addison-Wesley: Boston, 1990, p. 39.
\bibitem{hawk}  Hawking,  S. W. Black Hole Explosions?, {\em Nature}  {\bf 1974},  {\em 248}, 30-31.
\bibitem{revbennet} Bennet, C. H.  Logical Reversibility of Computation,  {\em IBM Journ. Res. and Dev.} {\bf 1973} {\em 17}, 525-532.  
\bibitem{fredtof}  Fredkin, E. and Toffoli, T. Conservative Logic, {\em Int. Journ. Theor. Phys.} {\bf  1982}  {\em 21}, 219-253.
\bibitem{landauer}  Landauer, R.  Irreversibility and Heat Generation in the Computing Process, {\em IBM Journ. Res. Dev.}  {\bf 1961}  {\em 4}, 183.
\bibitem{preskill}  Preskill, J.  Do Black Holes Destroy Information?, in {\em Proceedings of the International Symposium on Black Holes, Membranes, Wormholes and Superstrings},  Kalara, S.  and Nanopoulos, D.V. ,  Eds., World Scientific: Singapore, 1993, pp. 22-39. 
\bibitem{holotofft}   t'Hooft, G. Dimensional Reduction in Quantum Gravity,  arXiv:gr-qc/9310026v2.
\bibitem{holosuss}  Susskind, L.   The World as a Hologram,  {\em J. Math. Phys.}
  {\bf 1995}, {\em 36},  6377-6396.  
\bibitem{Schr35}  Schr\"odinger, E.  Discussion of Probability Relations between Separated Systems, {\em Proc. Camb. Phil. Soc.} {\bf 1935},  {\em 31}, 555-563.
\bibitem{statmec} Popescu, S.,  Short,  A. J., and Winter, A.  Entanglement and the Foundations of Statistical Mechanics, {\em Nature Physics} {\bf 2006},  {\em 2(11)}, 754-758.
\bibitem{preskillnotes}   Preskill, J.   Lecture notes on Quantum Computation, http://www.theory.caltech.edu/people/preskill/ph229/ 
\bibitem{watrous}  Watrous, J.  Quantum  Information and Computation Lecture Notes, https://cs.uwaterloo.ca/~watrous/lecture-notes.html  
\bibitem{wilde}  Wilde, M.    {\em From Classical to Quantum Shannon Theory}, http://arxiv.org/abs/1106.1445.
\bibitem{refframe}  Chiribella, G.,  D'Ariano, G. M., Perinotti, P.,  and Sacchi, M. F.    Efficient Use of Quantum Resources for the Transmission of a Reference Frame {\em Phys. Rev. Lett.}  {\bf 2004}    {\em 93}, 180503.
\bibitem{giuliolec} Chiribella, G.  Group Theoretic Structures in the Estimation of an Unknown Unitary Transformation   J{\em  Phys. Conf. Ser. }   {\bf 2011}    {\em 284}, 012001.     
\bibitem{davidovich}  Escher, B. M.,   de Matos Filho, R. L., and Davidovich, L. General Framework for Estimating the Ultimate Precision Limit in Noisy Quantum-Enhanced Metrology, {\em Nature Physics}  {\bf 2011} {\em 7}, 406. 
\bibitem{wootterssharing}  Wootters, W. K., Entanglement Sharing in Real-Vector-Space Quantum Theory,   arXiv:1007.1479.
\bibitem{hardycomp}  Hardy, L.     Quantum Gravity Computers: On the Theory of Computation with Indefinite Causal Structure   in \emph{Quantum Reality, Relativistic Causality, and Closing the Epistemic Circle: Essays in Honour of Abner Shimony}, Myrvold, W. C. and Christian, J.  eds., Springer,  2009.
\bibitem{switch}  Chiribella, G., DÕAriano, G. M. and Perinotti, P.  Beyond Causally-Ordered Quantum Computers, arXiv:0912.0195.
\bibitem{brukner}  Oreshkov, O., Costa, F. and Brukner , \v C.  Quantum Correlations with No Causal Order,  arXiv:1105.4464; {\em Nature Communication}, in press.
\bibitem{giulio} Chiribella, G.  Perfect Discrimination of No-Signalling Channels via Quantum Superposition of Causal Structures,  arXiv:1109.5154. 
\bibitem{programmable}  Colnaghi, T., D.Ariano, G. M., Perinotti, P., and Facchini, S. Quantum Computation with Probrammable Connections Between Gates, arXiv:1109.5987; {\em Phys. Lett. A}, in press.     
  \end{thebibliography}
%----------------------------------------------------------

\end{document}